\documentclass[journal, 10pt, twocolumn]{IEEEtran}

\usepackage{epsf}
\usepackage{graphics}
\usepackage{times}
\usepackage{graphicx}
\usepackage{amsmath}
\usepackage{amssymb}
\usepackage{epsfig}
\usepackage{subfigure}
\usepackage{array}
\usepackage{float}

\begin{document}

\title{Dither Signal Design for PAPR Reduction in OFDM-IM over a Rayleigh Fading Channel}

\author{Kee-Hoon Kim
\thanks{The author is with the School of Electronic and Electrical Engineering and IITC, Hankyong National University, Anseong 17579, South Korea (e-mail: keehk85@gmail.com)}}
\maketitle

\begin{abstract}
Orthogonal frequency division multiplexing with index modulation (OFDM-IM) is a novel scheme where the information bits are conveyed through the subcarrier activation pattern (SAP) and the symbols on the active subcarriers. Unfortunately, OFDM-IM inherits the high peak-to-average power ratio (PAPR) problem from the classical OFDM. The OFDM-IM signal with high PAPR induces in-band distortion and out-of-band radiation when it passes through high power amplifier (HPA).
There are attempts to reduce PAPR by adding dither signals in the idle subcarriers, where the dither signals can have various amplitude constraints according to the characteristic of the corresponding OFDM-IM subblock. But, there is no result for the specific amplitude constraint for the dither signals over a Rayleigh fading channel. In this letter, based on pairwise error probability (PEP) analysis, a specific constraint for the dither signals is derived over a Rayleigh fading channel.
\end{abstract}


\section{Introduction}

Orthogonal frequency division multiplexing with index modulation (OFDM-IM) \cite{bacsar2013orthogonal} is an emerging technique which is the application of the spatial modulation (SM) \cite{mesleh2008spatial} principle to the subcarriers in an OFDM system. In OFDM-IM, the subcarriers are partitioned into many subblocks and, in each subblock, the subcarriers have two states, active or idle. Then, OFDM-IM conveys information through not only modulated symbols but also the indices of the active subcarriers. For the same spectral efficiency, OFDM-IM is shown to have a superior bit error rate (BER) performance when compared to the classical OFDM \cite{bacsar2013orthogonal}. Also, OFDM-IM systems have a better energy efficiency compared to the classical OFDM \cite{zhao2012high}.

The authors in \cite{ishikawa2016subcarrier} pointed out that OFDM-IM inherits the high peak-to-average power ratio (PAPR) problem from the classical OFDM. It is known that the high PAPR induces in-band distortion and out-of-band radiation in consideration of nonlinear high power amplifier (HPA). Numerous PAPR reduction schemes have been researched for decades such as selected mapping (SLM), partial transmit sequence (PTS), and tone injection (TI) \cite{han2005overview}. To solve the high PAPR problem in OFDM-IM, we may borrow the PAPR reduction schemes designed for the classical OFDM. However, those methods are not efficient because they do not consider the unique characteristic of OFDM-IM structure. Specifically, they do not exploit the idle subcarriers in OFDM-IM \cite{zheng2017peak}.

Therefore, the authors in \cite{zheng2017peak} proposed a PAPR reduction method using the idle subcarriers in OFDM-IM. In specific, the scheme in \cite{zheng2017peak} introduces dither signals in the idle subcarriers for reducing PAPR of OFDM-IM signals. This is the first PAPR reduction method exploiting the special structure of OFDM-IM. This methodology is quite reasonable because the dither signals in the idle subcarriers do not considerably affect the error performance in high signal-to-noise ratio (SNR) region. This is because the symbol error event with diversity order of one dominates the system performance in the high SNR region and the dither signals cannot affect this error event. To suppress the harmful effect of the dither signals, they also proposed the equivalent amplitude constraint for the dither signals.

Besides, the previous work in \cite{kim2019papr} considered the fact that the amplitude characteristics of the subblocks are distinct when quadrature amplitude modulation (QAM) is employed in the active subcarriers. Therefore, a variable amplitude constraint for dither signals is proposed. As a result, the constraint in \cite{kim2019papr} gives the dither signals more freedom in average while maintaining good demodulation performance.
However, in \cite{kim2019papr}, the amplitude constraint is derived under the assumption of an additive white Gaussian noise (AWGN) channel without fading. Also, the low complexity power based detection algorithm is considered in the derivation, where the power based detection is less preferred in the recent literature because of its degraded performance.

In this letter, based on rigorous pairwise error probability (PEP) analysis, a new amplitude constraint for the dither signals is proposed over a Rayleigh fading channel, which is theoretically meaningful. By using the proposed amplitude constraint, dither signals can be well designed over a fading channel.

\section{System Model}

\subsection{OFDM-IM}

Consider an OFDM-IM system with $N$ subcarriers and total $m$ information bits. These $m$ information bits are divided into $g$ subblocks, where each subblock conveys $p$ bits, i.e., $m=pg$. Also, the $p$ bits in each subblock are mapped to one subblock of length $n$ in the frequency domain, where $n = N/g$.
The specific mapping procedure is that only $k$ out of $n$ subcarriers in each subblock are activated, where the subcarrier activation pattern (SAP) is determined by the first $p_1$ bits of the $p$ bits. Then, the $M$-ary modulated symbols on the $k$ active subcarriers are determined by the remaining $p_2=k \log_2M$ bits of the $p$ bits, i.e., $p = p_1 + p_2$. Of course, the idle subcarriers have zero values in the frequency domain \cite{bacsar2013orthogonal}.

We denote the set of the indices of the $k$ active subcarriers in the $\beta$-th OFDM-IM subblock, $\beta=0,1,\cdots,g-1$, as
\begin{equation}
I^{\beta} = \{i_{0}^\beta,i_{1}^\beta,\cdots,i_{k-1}^\beta\}
\end{equation}
with $i_{\gamma}^\beta \in \{0,1,\cdots,n-1\}$ for $\gamma = 0,1,\cdots,k-1$.
Also, we denote the set of $k$ modulated symbols as
\begin{equation}
S^{\beta} = \{S_{0}^\beta,S_{1}^\beta,\cdots,S_{k-1}^\beta\}
\end{equation}
where $S_{\gamma}^\beta \in \mathcal{S}$ and $\mathcal{S}$ is the $M$-ary signal constellation.

By considering $I^{\beta}$ and $S^{\beta}$, the $\beta$-th OFDM-IM subblock in frequency domain can be generated as
\begin{equation}
X^{\beta} = [X_0^{\beta}~X_1^{\beta}~\cdots~X_{n-1}^{\beta}]^T
\end{equation}
where $X_i^{\beta}$ is the $i$-th element of $X^{\beta}$ and $i=0,1,\cdots,n-1$.
After all of the $g$ subblocks $\{X^{\beta}\}_{\beta=0}^{g-1}$ are generated, they are concatenated into the symbol sequence $\mathbf{X}$ of length $N$ in frequency domain. To achieve the frequency diversity gain as much as possible, concatenation in an interleaved pattern is generally employed \cite{xiao2014ofdm}.
By the interleaved pattern, the elements in an OFDM-IM subblock can experience independent fading channels.

Then, to obtain the OFDM-IM signal sequence $\mathbf{x}$ in time domain, the symbol sequence $\mathbf{X}$ in frequency domain is processed by the inverse discrete Fourier transform (IDFT) as
\begin{equation}
\mathbf{x} = \mbox{IDFT}(\mathbf{X}).
\end{equation}
For transmission, cyclic prefix (CP) insertion and digital-to-analog (D/A) conversion are sequentially performed and then the PAPR of the resultant continuous-time OFDM-IM signal ${x}(t)$ is defined by
\begin{equation}
\mbox{PAPR}({x}(t))=\frac{\max_t|{x}(t)|^2}{\mathbb{E}[|{x}(t)|^2]}.
\end{equation}
Practically, to capture PAPR of the continuous-time OFDM-IM signal, four-times oversampling of $\mathbf{x}$ is used.

The receiver firstly detects the SAP, called index demodulation in this letter, and demodulates the $M$-ary modulated symbols in the active subcarriers. There have been several detection algorithms for OFDM-IM systems \cite{bacsar2013orthogonal, siddiq2016low}. For low computational complexity, a power based detection algorithm can be employed, but it degrades the error performance of OFDM-IM. In \cite{zheng2015low, zhang2017dual}, it is known that the optimal maximum likelihood (ML) detection can be implemented with low complexity and thus it is preferred to employ the ML detection.

\section{PAPR Reduction Using Dither Signals}
\subsection{PAPR Reduction Using Dither Signals of an Equivalent Amplitude Constraint \cite{zheng2017peak}}

In OFDM-IM, there are two types of error events, an index demodulation error event and a symbol error event. The former is the error event when the SAP is incorrectly detected and the latter is the error event when the modulated symbols in the active subcarriers are incorrectly detected though the SAP is correctly detected.
As described in \cite{bacsar2013orthogonal}, the symbol error event has the diversity order of one and the index demodulation error event has the diversity order of two. By virtue of the frequency diversity gain, the index demodulation error event occurs less frequently than the symbol error event in high SNR region. That is, the symbol error event dominates the overall error performance of the OFDM-IM system in high SNR region.
Therefore, in \cite{zheng2017peak} there is an attempt to reduce PAPR by inserting the dither signals in the idle subcarriers. Clearly, this dither signal does not affect the symbol error event and only degrades the index demodulation error performance.
Also, the authors in \cite{zheng2017peak} introduces an equivalent amplitude constraint for the dither signals in the idle subcarriers.

Specifically, the dither signal added in the $\beta$-th OFDM-IM subblock is
\begin{equation}
D^\beta = [D_0^\beta~D_1^\beta~\cdots~D_{n-1}^\beta]^T
\end{equation}
and the constraint in \cite{zheng2017peak} with a hyperparameter $R$ is
\begin{equation}\label{eq:legacy}
D_{i}^{\beta}=\left\{
       \begin{array}{ll}
         |D_{i}^{\beta}|<R, & \hbox{$i\in (I^{\beta})^c$} \\
         0, & \hbox{$i\in I^{\beta}$}
       \end{array}
     \right.
\end{equation}
where $(I^{\beta})^c$ is the complement set of $I^{\beta}$, i.e., indices of the idle subcarriers.
Then, $X^\beta+D^\beta$ becomes a new $\beta$-th OFDM-IM subblock with dither signals. Under the constraint in (\ref{eq:legacy}), the values of $D^\beta$ are determined in order to reduce PAPR of $x(t)$. In \cite{zheng2017peak}, convex programming is used, but iterative clipping and filtering with trimming the dither signals to satisfy the constraint in (\ref{eq:legacy}) without phase change can be alternatively used for low complexity.

\subsection{PAPR Reduction with Dither Signals of a Variable Amplitude Constraint \cite{kim2019papr}}

Meanwhile, in \cite{kim2019papr}, a variable amplitude constraint for dither signals is proposed. This is motivated from the fact that the OFDM-IM subblocks have different robustness against channel noise if higher modulation than 16-QAM is employed. Using this, the amplitude constraint of dither signals can be varied for subblocks.
Let us briefly review the work in \cite{kim2019papr}.

First, for the $\beta$-th OFDM-IM subblock $X^{\beta}$, we define $A^\beta$ as
\begin{equation}
A^\beta = \min(|S_{0}^\beta|,|S_{1}^\beta|,\cdots,|S_{k-1}^\beta|).
\end{equation}
For ease of understanding, we assume that 16-QAM is employed with the signal constellation $\{\pm1\pm j1, \pm1\pm j3, \pm3\pm j1, \pm3\pm j3\}$ from now on. (Clearly, the scheme in \cite{kim2019papr} can be easily described for higher modulations such as 64-QAM or 256-QAM.) Then, $A^\beta$ can be one of $\{\sqrt{2},\sqrt{10},\sqrt{18}\}$.

Second, assuming an AWGN channel with noise power $N_0$, the PEP for the $\beta$-th subblock is
\begin{equation}\label{eq:PEPawgn}
P(X\rightarrow \hat{X}) = Q\left( \frac{\|X-\hat{X}\|}{\sqrt{2N_0}} \right)
\end{equation}
where, without loss of generality, we omit $\beta$ for simplicity and $\hat{X}=[\hat{X}_0~\hat{X}_1~\cdots~\hat{X}_{n-1}]^T$.
Also, $Q(\cdot)$ means the Q-function, the tail distribution function of the standard normal distribution.
It is clear that the PEP in (\ref{eq:PEPawgn}) depends on the Euclidean distance between $X$ and $\hat{X}$, $\|X-\hat{X}\|$.

Third, we focus on the index demodulation error event because the dither signals do not affect the symbol error event as we mentioned. Then, the fundamental index demodulation error in the $\beta$-th subblock is the case when the $u$-th subcarrier is detected as active and the $v$-th subcarrier is detected as idle when actually the opposite is true. The shortest Euclidean distance inducing this fundamental index demodulation error is
\begin{equation}
\min\sqrt{|\hat{X}_u|^2 + |X_v|^2} = \sqrt{2 + (A^\beta)^2}
\end{equation}
where the index demodulation error performance of the $\beta$-th subblock is dominated by this metric.
That is, as $A_\beta$ becomes larger, the robustness of the $\beta$-th OFDM-IM subblock against the index demodulation error increases. Therefore, dither signals with large amplitude may be inserted if $A_\beta$ is large.

Using this fact, in \cite{kim2019papr}, the dither signals can have different amplitudes according to $A^{\beta}$ as
\begin{equation}
D_{i}^{\beta}=\left\{
        \begin{array}{ll}
          |D_{i}^{\beta}|<R_0, & \hbox{$i\in (I^{\beta})^c$ and $A^\beta=\sqrt{2}$} \\
          |D_{i}^{\beta}|<R_1, & \hbox{$i\in (I^{\beta})^c$ and $A^\beta=\sqrt{10}$} \\
          |D_{i}^{\beta}|<R_2, & \hbox{$i\in (I^{\beta})^c$ and $A^\beta=\sqrt{18}$} \\
          0, & \hbox{$i\in I_{\beta}$}.
        \end{array}
      \right.
\end{equation}
Also, the constraint proposed in \cite{kim2019papr} is
\begin{equation}\label{eq:cri}
\sqrt{2} - R_0 = \sqrt{10} - R_1 = \sqrt{18} - R_2
\end{equation}
where they assume the AWGN channels and using the power based detection algorithm at the receiver.

One example of the possible candidates is $R_0=0.2$, $R_1\simeq1.9$, and $R_2\simeq3$, which gives the dither signals more freedom than the equivalent amplitude constraint with $R=0.2$ in (\ref{eq:legacy}). However, the constraint in (\ref{eq:cri}) is inappropriate for frequency selective fading channels because the high values of $R_1$ and $R_2$ destroy the diversity.

\section{Proposed Dither Signals Design over a Rayleigh Fading Channel}

In this section, we propose an amplitude constraint of the dither signals with consideration of a Rayleigh fading channel.
First, we denote the channel frequency response (CFR) of the $i$-th element in the $\beta$-th OFDM-IM subblock as $H_{i}$ (with omission of $\beta$) and the CFR matrix is denoted as $H = \mathrm{diag}(H_0,H_1,\cdots,H_{n-1})$. Note that $H_0,H_1,\cdots,H_{n-1}$ are approximately independent because we use concatenation in an interleaved pattern. (This is a valid assumption unless we use channels with less than $n$ taps in time domain.)
For a given matrix $H$, the conditional PEP of the $\beta$-th OFDM-IM subblock with the dither signal $D$ is
\begin{align}
&P( X + D \rightarrow \hat{X}|H)\\
&= P(\| Y-H\hat{X}\|^2 <\| Y-H{X}\|^2|H)\\
&=P(\| H(X+D)+Z-H\hat{X}\|^2 <\| HD + Z\|^2|H)\\
&=P(2\cdot\mathrm{Re}\{(HD+Z)^H H(X-\hat{X})\} <-\| H(X-\hat{X})\|^2|H)\label{eq:PEP0}
\end{align}
where $Y = H(X+D)+Z$ is the $\beta$-th received OFDM-IM subblock and $Z$ is AWGN with $Z_{i}\sim \mathcal{CN}(0,N_0)$.

Since the dither signals cannot affect the symbol error events, let us consider the fundamental index demodulation error case that the $u$-th subcarrier is detected as active and the $v$-th subcarrier is detected as idle when actually the opposite is true. Then, (\ref{eq:PEP0}) becomes
\begin{align}
&P( X + D \rightarrow \hat{X}|H)\\
&=P(\mathrm{Re}\{-|H_u|^2 D_u^*\hat{X}_u - H_u Z_u^* \hat{X}_u + H_vZ_v^*  X_v\}\\
&~~~~~~<-\frac{1}{2}( |H_u\hat{X}_u|^2 + |H_vX_v|^2 )|H)\label{eq:PEP1}
\end{align}
where $D_u$ is the dither signal in the $u$-th idle subcarrier in $X$.
If we consider the dither signal $D_u$ satisfying
\begin{equation}
|\mathrm{Re}\{D_u\}|+|\mathrm{Im}\{D_u\}| = \sqrt{2}R
\end{equation}
then the left hand side (LHS) term in (\ref{eq:PEP1}) becomes
\begin{equation}\label{eq:LHSterm}
-R|H_u|^2|\hat{X}_u|+\mathrm{Re}\{- H_u Z_u^* \hat{X}_u + H_vZ_v^*  X_v\}
\end{equation}
under the assumption of $|\mathrm{Re}\{\hat{X}_u\}| = |\mathrm{Im}\{\hat{X}_u\}|$ and $\hat{X}_u$ and $D_u$ lie on the same quadrant. (This assumption is valid because we will consider the weakness case inducing the index demodulation error.)
Clearly, (\ref{eq:LHSterm}) is Gaussian distributed as
\begin{equation}
\mathcal{N}(-R|H_u|^2|\hat{X}_u|,\frac{N_0}{2}(|H_u\hat{X}_u|^2 + |H_vX_v|^2)).
\end{equation}

\begin{figure}[htbp]
\centering
\includegraphics[width=.8\linewidth]{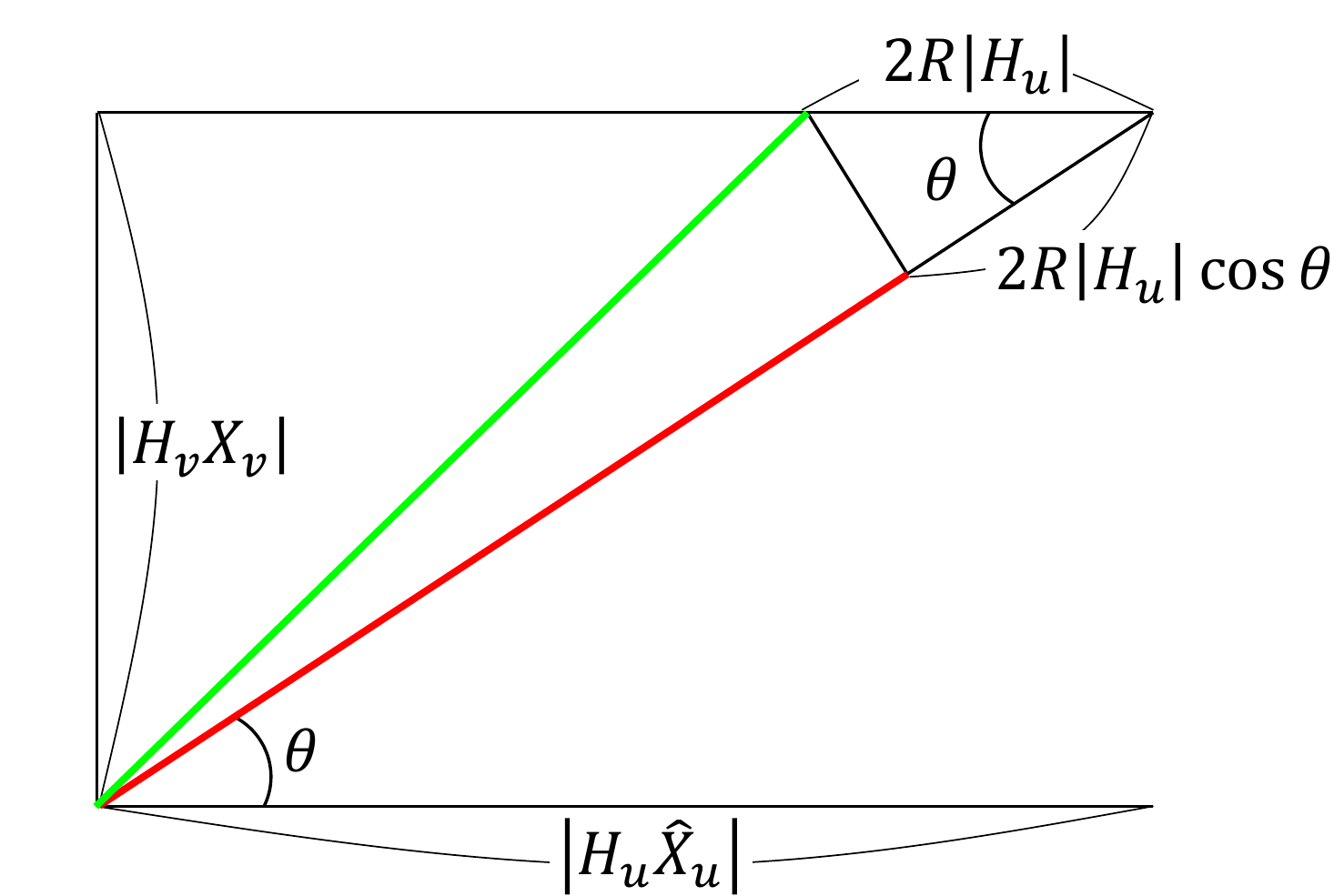}
\caption{A geometrical representation of $\sqrt{|H_u\hat{X}_u|^2 + |H_vX_v|^2}- \frac{2R|H_u||H_u\hat{X}_u|}{\sqrt{|H_u\hat{X}_u|^2 + |H_vX_v|^2}}$ by the led line. Note that $\frac{|H_u\hat{X}_u|}{\sqrt{|H_u\hat{X}_u|^2 + |H_vX_v|^2}}$ is $\cos \theta$.}
\label{fig:geo}
\end{figure}

Then, (\ref{eq:PEP1}) becomes
\begin{align}
&P( X + D \rightarrow \hat{X}|H)\\
&=Q\left(\frac{|H_u\hat{X}_u|^2 + |H_vX_v|^2 - 2R|H_u|^2|\hat{X}_u|}{\sqrt{2N_0}\sqrt{|H_u\hat{X}_u|^2 + |H_vX_v|^2}}\right)\\
&=Q\left(\frac{1}{\sqrt{2N_0}}\left(\sqrt{|H_u\hat{X}_u|^2 + |H_vX_v|^2}- \frac{2R|H_u||H_u\hat{X}_u|}{\sqrt{|H_u\hat{X}_u|^2 + |H_vX_v|^2}}\right)\right)\label{eq:geo}
\end{align}
where the term $\sqrt{|H_u\hat{X}_u|^2 + |H_vX_v|^2}- \frac{2R|H_u||H_u\hat{X}_u|}{\sqrt{|H_u\hat{X}_u|^2 + |H_vX_v|^2}}$ in (\ref{eq:geo}) can be depicted as the red line
in Fig. \ref{fig:geo}.

As seen in Fig. \ref{fig:geo}, the length of the red line can be alternatively approximated as the length of the green line. This is approximation is reasonable because $|H_u\hat{X}_u|$ and $|H_vX_v|$ are similar based on the fact that $\mathbb{E}[|H_u\hat{X}_u|]=\mathbb{E}[|H_vX_v|]$ and the value of $R$ is relatively small compared to $|\hat{X}_u|$.
The length of the green line is $\sqrt{|H_u(|\hat{X}_u|-2R)|^2+|H_vX_v|^2}$ in Fig. \ref{fig:geo} and thus (\ref{eq:geo}) becomes
\begin{equation}\label{eq:PEP2}
P( X + D \rightarrow \hat{X}|H)\simeq Q\left(\frac{\sqrt{|H_u(|\hat{X}_u|-2R)|^2+|H_vX_v|^2}}{\sqrt{2N_0}}\right).
\end{equation}

Meanwhile, in \cite{simon2005digital, van2018spread}, it is known that the unconditional PEP with independent Rayleigh channel frequency responses $H_i \sim \mathcal{CN} (0, 1)$ in high SNR region is expressed as
\begin{align}
&P( X  \rightarrow \hat{X})\\
&= \int P( X  \rightarrow \hat{X}|H)p(H)dH\\
&= \int Q\left(\frac{\| H(X-\hat{X})\|}{\sqrt{2N_0}}  \right)p(H)dH\\
&\simeq \frac{(4N_0)^{\Gamma_{X,\hat{X}}}}{2\Pi_{i\in \mathcal{G}_{X,\hat{X}}} \eta_i }\label{eq:uncPEP}
\end{align}
where $p(H)$ is the probability distribution function (pdf) of $H$, $\eta_i$ is the $i$-th element of $|X-\hat{X}|^2$, $\mathcal{G}_{X,\hat{X}} = \{i|\eta_i\neq 0 \}$, and $\Gamma_{X,\hat{X}} = |\mathcal{G}_{X,\hat{X}}|$.
By combining (\ref{eq:PEP2}) and (\ref{eq:uncPEP}), we have
\begin{equation}
P( X + D \rightarrow \hat{X})\simeq \frac{(4N_0)^{2}}{2 (|\hat{X}_u|-2R)^2\cdot |X_v|^2}.
\end{equation}
That is, the unconditional PEP $P( X + D \rightarrow \hat{X})$ depends on the metric as
\begin{equation}
(|\hat{X}_u|-2R)\cdot |{X}_v|.
\end{equation}
Then, the weakness case inducing the fundamental index demodulation error in the $\beta$-th subblock depends on the metric as
\begin{equation}\label{eq:metric}
\min(|\hat{X}_u|-2R)\cdot |X_v| = (\sqrt{2}-2R)\cdot A^{\beta}.
\end{equation}
Clearly, the index demodulation error performance of the $\beta$-th OFDM-IM subblock is dominated by the value of (\ref{eq:metric}).

As in the scheme in \cite{kim2019papr}, the proposed scheme is valid if QAM modulation is considered. If 16-QAM is employed with the signal constellation $\{\pm1\pm j1, \pm1\pm j3, \pm3\pm j1, \pm3\pm j3\}$, the proposed variable amplitude constraint is
\begin{align}
&D_{i}^{\beta}\\
&=\left\{
        \begin{array}{ll}
          |\mathrm{Re}\{D_{i}^{\beta}\}| + |\mathrm{Im}\{D_{i}^{\beta}\}|<\sqrt{2}R_0, & \hbox{$i\in (I^{\beta})^c$, $A_\beta=\sqrt{2}$} \\
          |\mathrm{Re}\{D_{i}^{\beta}\}| + |\mathrm{Im}\{D_{i}^{\beta}\}|<\sqrt{2}R_1, & \hbox{$i\in (I^{\beta})^c$, $A_\beta=\sqrt{10}$} \\
          |\mathrm{Re}\{D_{i}^{\beta}\}| + |\mathrm{Im}\{D_{i}^{\beta}\}|<\sqrt{2}R_2, & \hbox{$i\in (I^{\beta})^c$, $A_\beta=\sqrt{18}$} \\
          0, & \hbox{$i\in I_{\beta}$}
        \end{array}
      \right.
\end{align}
with
\begin{equation}\label{eq:const}
(\sqrt{2}-2R_0)\cdot\sqrt{2} = (\sqrt{2}-2R_{1})\cdot\sqrt{10} =  (\sqrt{2}-2R_2)\cdot\sqrt{18}
\end{equation}
which leads to the same value of (\ref{eq:metric}) for all subblocks. (Since the bottleneck of the OFDM-IM error performance relates to the smallest value of (\ref{eq:metric}) for all $g$ subblocks, this strategy could be the optimal solution.)
It is straightforward to generate the constraints for other modulations.

Table \ref{tab:ex} shows several examples of the values of $R_0$, $R_1$, and $R_2$ satisfying the proposed amplitude constraint in (\ref{eq:const}). We remark that the values of $R_1$ and $R_2$ are suppressed because of frequency selective fading channels, compared to the values from (\ref{eq:cri}).
\begin{table}
  \centering
\caption{Examples of $R_0$, $R_1$, and $R_2$ values satisfying (\ref{eq:const})}\label{tab:ex}
  \begin{tabular}{|c|c|c|}
    \hline
    $R_0$ & $R_1$ & $R_2$ \\ \hline
    0.1 & 0.435 & 0.504 \\ \hline
    0.2 & 0.480 & 0.538 \\ \hline
    0.3 & 0.525 & 0.571 \\ \hline
    0.4 & 0.569 & 0.604 \\ \hline
    0.5 & 0.614 & 0.638 \\
    \hline
  \end{tabular}
\end{table}

\section{Simulation Results}
For modulating the symbols in the active subcarriers, 16-QAM is used. Also, we use $N=128$, $n=4$, and $k=2$. The OFDM-IM subblocks are concatenated in an interleaved pattern and dither signals are generated by five times iterative clipping and filtering with trimming for all schemes.
To capture PAPR of the continuous-time OFDM-IM signal, four-times oversampling is used for the clipping and filtering. At the receiver, the optimal ML detection with low computational complexity is employed.

\begin{figure}[htbp]
\centering
\includegraphics[width=.95\linewidth]{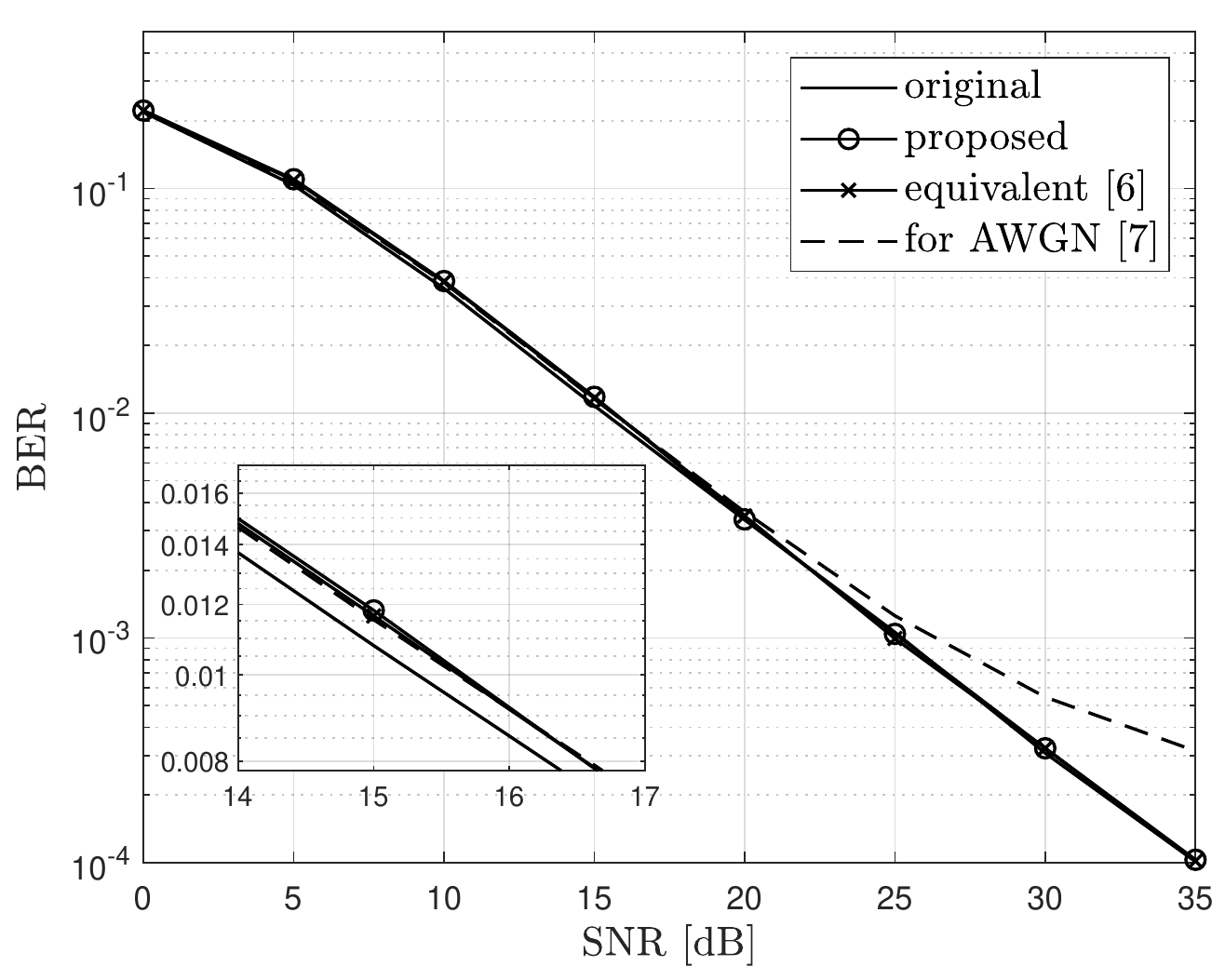}
\caption{BER performance of the OFDM-IM signals with different amplitude constraints.}
\label{fig:BER}
\end{figure}

Fig. \ref{fig:BER} shows the BER performance over a Rayleigh fading channel with eight channel taps.
The SNR means the average energy per bit over $N_0$. For fair comparison, both symbols in the active subcarriers and dither signals in the idle subcarriers are considered when we calculate the average energy per bit.
The meaning of four labels in the legend is as follows; the original OFDM-IM signal without clipping; the OFDM-IM signal under the proposed amplitude constraint, $R_0=0.5$, $R_1=0.614$, and $R_2=0.638$; the OFDM-IM signal under the equivalent amplitude constraint in \cite{zheng2017peak}, $R=0.5$; the OFDM-IM signal under the constraint in \cite{kim2019papr}, $R_0=0.2$, $R_1=1.9$, and $R_2=3$, which is originally designed for AWGN channels. In comparison, we do not consider the traditional PAPR reduction schemes such as SLM, PTS, and TI in \cite{han2005overview} because the PAPR reduction schemes exploiting the idle subcarriers do not exist.

As we expected, the proposed constraint and the equivalent amplitude constraint in \cite{zheng2017peak} have almost the same BER performance because the two constraints induce the same PEP for the bottleneck of the OFDM-IM error performance. In specific, they have the same smallest value of (\ref{eq:metric}) for all $g$ subblocks.
The constraint originally designed for AWGN channels in \cite{kim2019papr} shows the error floor over 25 dB SNR point because the large values of $R_1$ and $R_2$ destroy the diversity gain over a fading channel.

\begin{figure}[htbp]
\centering
\includegraphics[width=.95\linewidth]{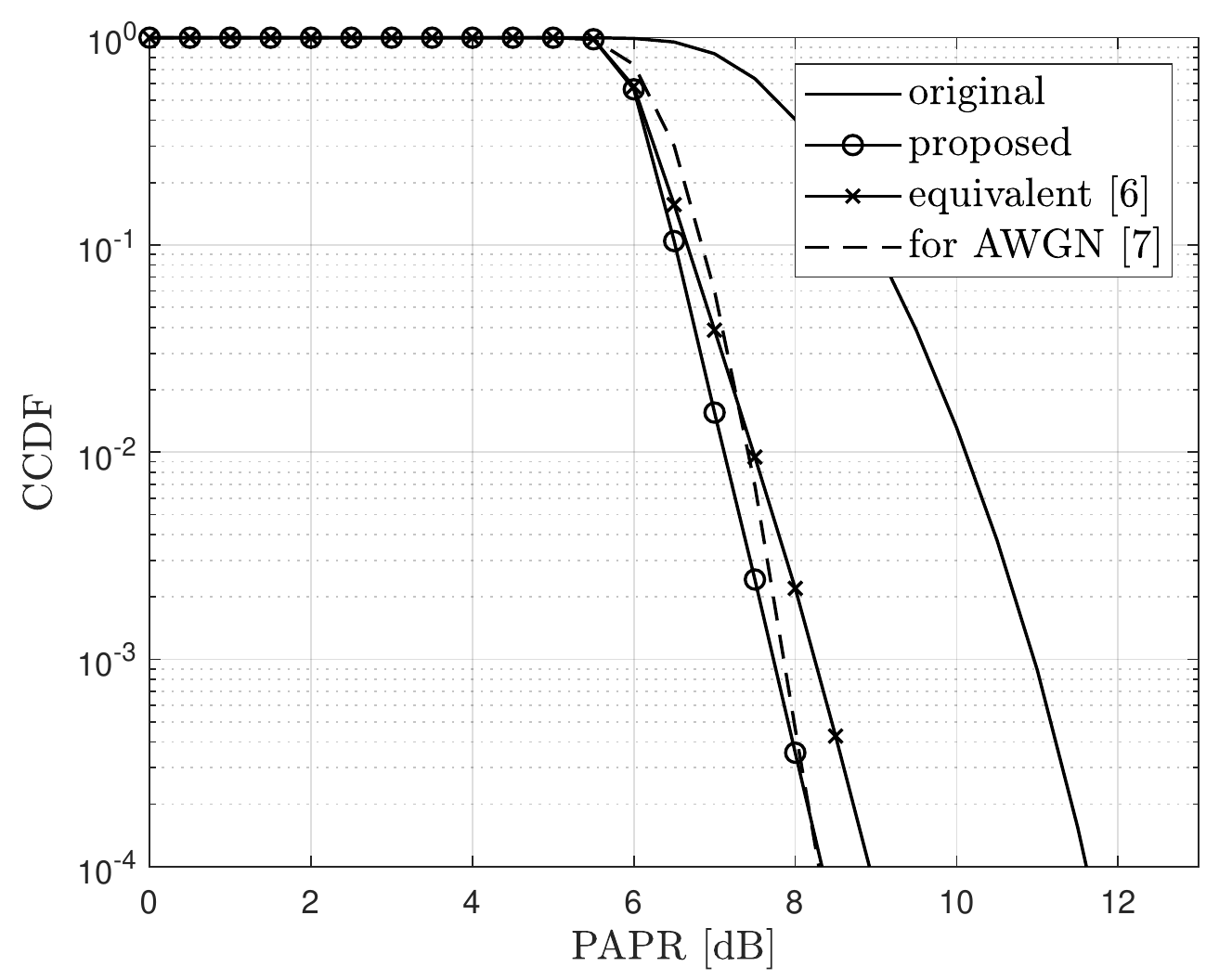}
\caption{PAPR reduction performance of the OFDM-IM signals with different amplitude constraints. Four-times oversampling is used.}
\label{fig:PAPR}
\end{figure}

Fig. \ref{fig:PAPR} shows the PAPR reduction performance of the four cases. The ordinate is the complementary cumulative distribution function (CCDF) of the PAPR. Using the proposed amplitude constraint can reduce much PAPR than the scheme in \cite{zheng2017peak} using the equivalent amplitude constraint. This result mainly comes from the fact that the proposed constraint has larger freedom of dither signals. Although the PAPR reduction performance gap between the two schemes is small, the specific constraint based on the rigorous analysis is theoretically meaningful to design the dither signals in OFDM-IM systems.

\section{Conclusion}
In this letter, we proposed the variable amplitude constraint for the dither signals for PAPR reduction in OFDM-IM systems over a Rayleigh fading channel, which is based on the rigorous PEP analysis. By using the proposed variable amplitude constraint, the PAPR reduction performance can be increased compared to the conventional case using the equivalent amplitude constraint with almost the same BER performance.

\bibliographystyle{IEEEtran}
\bibliography{biblio,IEEEfull}

\end{document}